\documentclass[sigconf,hdvipdfm,dvipsnames]{acmart}

\usepackage{graphicx}
\graphicspath{{./}}
\usepackage{subfigure}

\usepackage{url}

\usepackage{balance}

\usepackage{booktabs,multirow}

\usepackage{amsmath}

\usepackage{xcolor}

\copyrightyear{2017}
\acmYear{2017}
\setcopyright{acmlicensed}
\acmConference{SoCC '17}{September 24--27, 2017}{Santa Clara, CA, USA}
\acmPrice{15.00}
\acmDOI{10.1145/3127479.3128605}
\acmISBN{978-1-4503-5028-0/17/09}

\begin{document}
\title[BestConfig: Tapping the Performance Potential of Systems via ACT]{BestConfig: Tapping the Performance Potential of Systems via Automatic Configuration Tuning}
\titlenote{Yuqing Zhu is the corresponding author.\vspace{-9pt}}

\author{Yuqing Zhu, Jianxun Liu, Mengying Guo, Yungang Bao, Wenlong Ma,\\Zhuoyue Liu, Kunpeng Song, Yingchun Yang}
\affiliation{%
  \institution{Institute of Computing Technology, Chinese Academy of Sciences\\Huawei}
  \city{\{zhuyuqing,liujianxun,guomengying,baoyg,mawenlong\}@ict.ac.cn\\\{liuzhuoyue,songkunpeng,yingchun.yang\}@huawei.com}\vspace{18pt}
}

%

\renewcommand{\shortauthors}{Y. Zhu et al.}

\begin{abstract}
An ever increasing number of configuration parameters are provided to system users. But many users have used one configuration setting across different workloads, leaving untapped the performance potential of systems. A good configuration setting can greatly improve the performance of a deployed system under certain workloads. But with tens or hundreds of parameters, it becomes a highly costly task to decide which configuration setting leads to the best performance. While such task requires the strong expertise in both the system and the application, users commonly lack such expertise.

To help users tap the performance potential of systems, we present BestConfig, a system for automatically finding a best configuration setting within a resource limit for a deployed system under a given application workload. BestConfig is designed with an extensible architecture to automate the configuration tuning for general systems. To tune system configurations within a resource limit, we propose the divide-and-diverge sampling method and the recursive bound-and-search algorithm. BestConfig can improve the throughput of Tomcat by 75\%, that of Cassandra by 63\%, that of MySQL by 430\%, and reduce the running time of Hive join job by about 50\% and that of Spark join job by about 80\%, solely by configuration adjustment.\vspace{-3pt}
\end{abstract}


\keywords{automatic configuration tuning, ACT, performance optimization\vspace{-15pt}}

\maketitle

\vspace{-12pt}\section{Introduction}

More and more configuration parameters are provided to users, as systems in the cloud aim to support a wide variety of use cases~\cite{knobs}. For example, Hadoop~\cite{hadoop}, a popular big data processing system in the cloud, has more than 180 configuration parameters. The large number of configuration parameters lead to an ever-increasing complexity of configuration issues that overwhelms users, developers and administrators. This complexity can result in configuration errors~\cite{gmailconfig,amazonconfig,msconfig,fbconfig}. It can also result in unsatisfactory performances under atypical application workloads~\cite{ituned,RLweb,smarthillclimbing,starfish}. In fact, configuration settings have strong impacts on the system performance~\cite{tuneSpark,tuneHadoop,tuneCassandra,tuneMysql}. To tap the performance potential of a system, system users need to find an appropriate configuration setting through configuration tuning.

\begin{figure*}
  \centering
  \subfigure[MySQL under uniform reads]{
    \label{fig:mysql} 
    \includegraphics[width=0.3\textwidth,height=103pt]{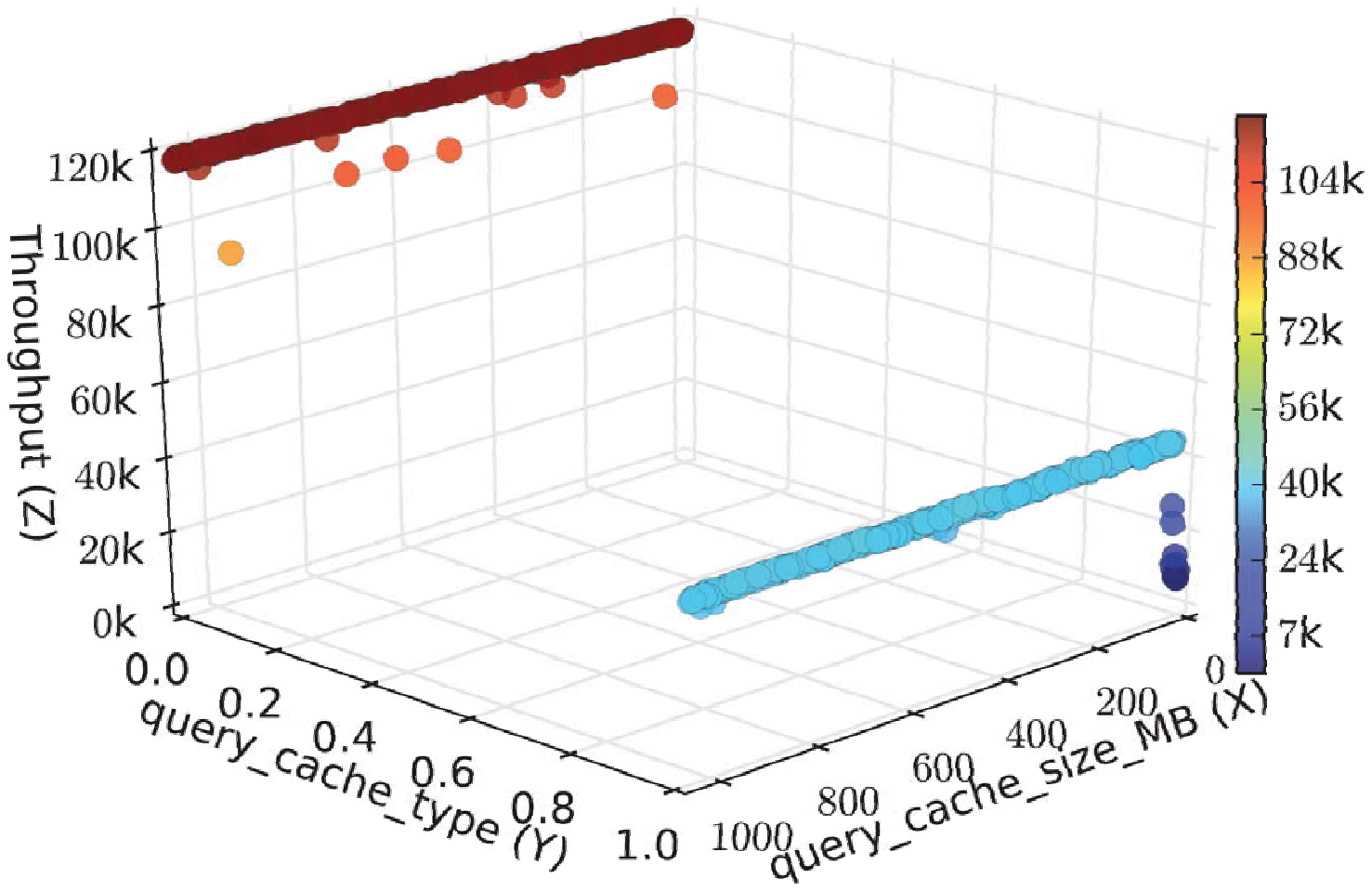}}%
  \subfigure[Tomcat under webpage navigation workload]{
    \label{fig:tomcat} 
    \includegraphics[width=0.34\textwidth,height=105pt]{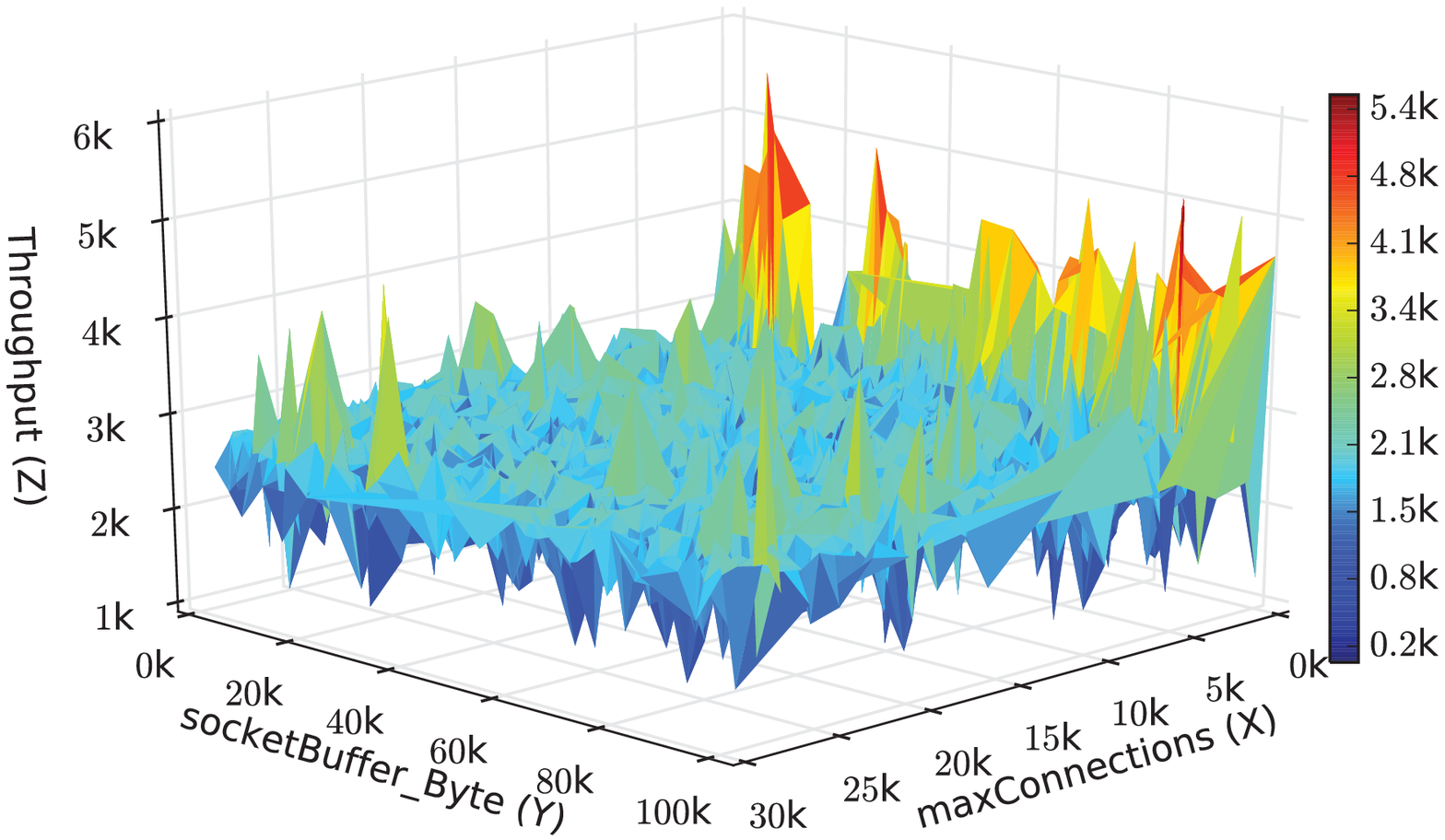}}
    \subfigure[Spark under HiBench-KMeans workload]{
    \label{fig:spark} 
    \includegraphics[width=0.32\textwidth,height=105pt]{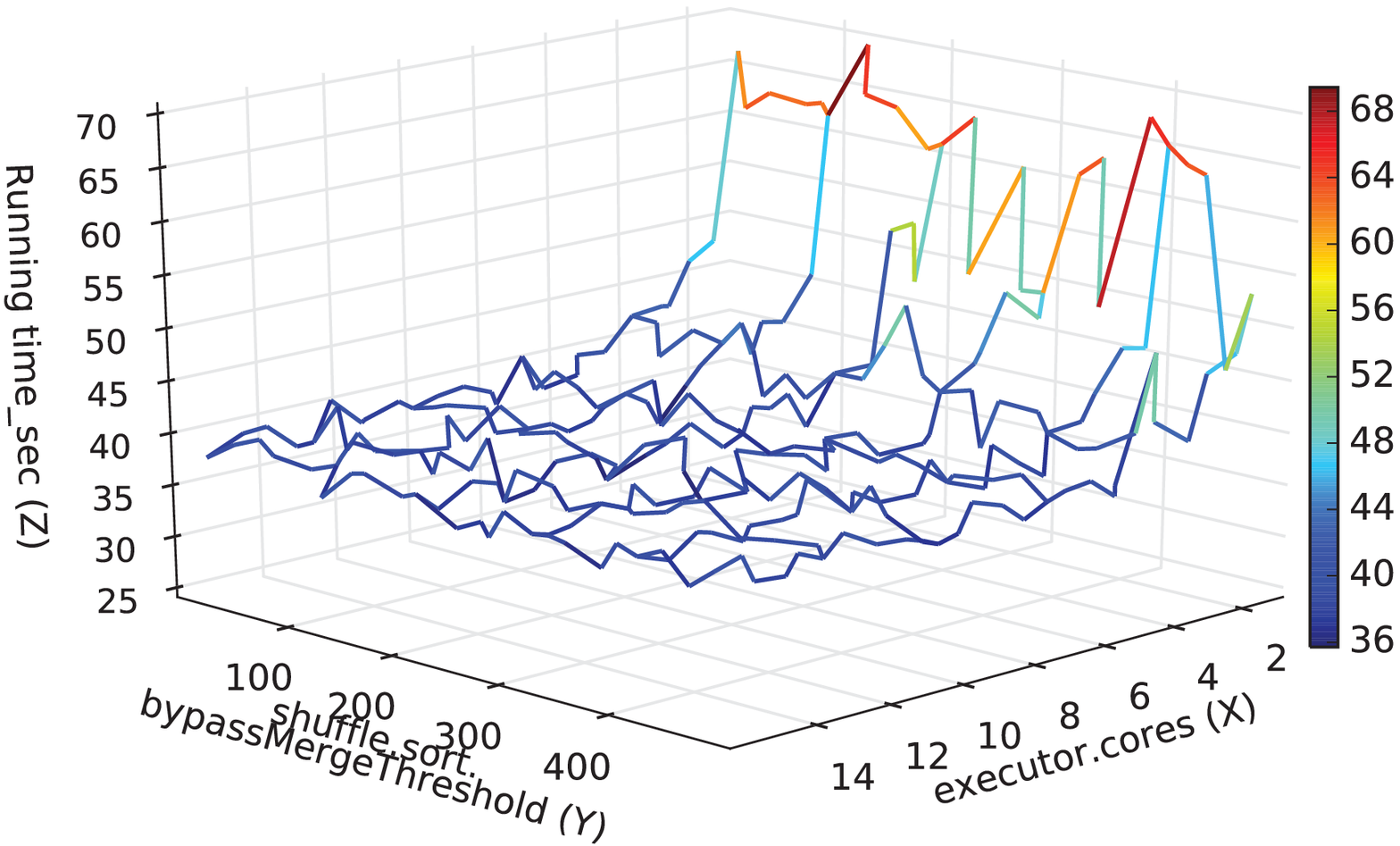}}
  \vspace{-12pt}\caption{\underline{Diverging} performance surfaces of \emph{MySQL}, \emph{Tomcat} and \emph{Spark}. (Best view in color)}\vspace{-9pt}
  \label{fig:perfFunc} 
\end{figure*}
A good configuration setting can greatly improve the system performance. For instance, changing the \emph{query\_cache\_type} parameter of MySQL from zero to one can result in more than \textbf{11 times} performance gain for an application workload, as shown in Figure~\ref{fig:mysql}. This performance gain can be significant if the workload is recurring on a daily base---this is very likely, especially for systems like databases or web servers. Nevertheless, configuration tuning for general systems is difficult due to the following three matters.\vspace{6pt}\\
\textbf{Variety}. Systems for configuration tuning can be data analytic systems like Hadoop~\cite{hadoop} and Spark~\cite{spark}, database systems like MySQL~\cite{mysql}, or web servers like Tomcat~\cite{tomcat}. Various deployments for a system are possible in the cloud. Performance goals concerning users can be throughput, latency, running time, etc. Among the variety of performance goals, some need to be maximized, while some minimized. The configuration tuning process must also take the application workload into account, and there are a variety of possible workloads. Furthermore, various combinations of systems, performance goals and workloads are possible.\\
\textbf{Complexity}. Given different performance goals and applied different workloads, a deployed system has different performance surfaces for a given set of configuration parameters. Different systems can have highly diverse and complex performance surfaces. Take Figure~\ref{fig:mysql}, \ref{fig:tomcat} and \ref{fig:spark} for example, MySQL has no performance surface but only two lines, while Tomcat has a bumpy performance surface and Spark has a relatively smooth performance surface. Previously, an unexpected performance surface is reported for PostgreSQL~\cite{ituned}, which even costs system developers months of efforts to reason about the underlying interactions.\\
\textbf{Overhead}. Configuration tuning involves solving a problem with a high-dimensional parameter space, thus a large sample set is commonly needed to find a solution~\cite{paraTuning}. However, collecting a large set of performance-configuration samples is impractical for configuration tuning. As no performance simulator exists for general systems, the samples can only be generated through real tests on the deployed system. Hence, configuration tuning must restrain the overhead of sample collection. Besides, the time overhead of the optimization process must also be considered.

Existing solutions do not fully address all the above challenges. Though sporadic proposals are found on automatically suggesting configuration settings for Web servers~\cite{smarthillclimbing,rrs,RLweb}, databases~\cite{ituned} and Hadoop~\cite{starfish,aloja} respectively, these solutions are not generally applicable to the variety of systems in the cloud. A few statistical or machine learning models are proposed for distributed systems~\cite{tkdeConfTune,resSurf}, but these models are not applicable to the complicated cases as shown from Figure~\ref{fig:mysql} to \ref{fig:spark}. Configuration tuning is related to the problem of optimizing the performance for systems with high-dimensional parameters~\cite{paraTuningBook}, but previous research typically studies the problem based on simulations~\cite{paraTuning}; the overhead aspect is rarely considered to the extent as required by configuration tuning for general systems.

In this paper, we present BestConfig---an automatic configuration tuning system that can optimize performance goals for general systems in the cloud by adjusting configuration parameters and that can recommend the best configuration setting found within a given resource limit. A typical resource limit is the number of tests allowed for configuration tuning. To address the resource limit challenge, BestConfig adopts an effective sampling method with wide space coverage and this coverage will be improved as more resources are provided. With the variety of systems and workloads, as well as the complexity of their interactions, it is impossible to build a useful performance model on a limited number of samples. Hence, BestConfig adopts a search-based optimization algorithm and exploits the general properties of performance models. To facilitate the usage with the variety of deployed systems and workloads, we design for BestConfig a software architecture that has loosely coupled but extensible components and that adopts a sample-test-optimize process in closed loop.

In the evaluation with extensive experiments, BestConfig can improve the throughput of Tomcat by 75\%, that of Cassandra~\cite{cassandra} by 63\%, that of MySQL by 430\%, and reduce the running time of Hive-over-Hadoop~\cite{hive} join job by about 50\% and that of Spark join job by about 80\%, as compared to the default configuration setting, simply by adjusting configuration settings.

In sum, this paper makes the following contributions.\vspace{-3pt}
\begin{itemize}
  \item To the best of our knowledge, we are the first to propose and the first to implement an automatic configuration tuning system for general systems. And, our system successfully automates the configuration tuning for six systems widely deployed in the cloud.
\item We propose an architecture ($\S$\ref{sec:arch}) that can be easily plugged in with general systems and any known system tests. It also enables the easy testing of other configuration tuning algorithms.
 \item We propose the divide-and-diverge sampling method ($\S$\ref{sec:DDS}) and the recursive-bound-and-search method ($\S$\ref{sec:RBS}) to enable configuration tuning for general systems within a resource limit.
 \item We demonstrate the feasibility and the benefits of BestConfig through extensive experiments ($\S$\ref{sec:eval}), while refusing the possibility of using common model-based methods such as linear or smooth prediction models for general systems ($\S$\ref{sec:model}).
  \item We have applied BestConfig to a real use case ($\S$\ref{sec:case}) showing that, even when a cloud deployment of Tomcat has a full resource consumption rate, BestConfig can still improve the system performance solely by configuration tuning.
\end{itemize}
\section{Background and Motivation}%
\label{sec:problem}

In this section, we describe the background and the motivation of automatic configuration tuning for general systems. We also analyze the challenges in solving this problem.

\subsection{Background}

\begin{figure}[!t]
      \centering
      \includegraphics[width=0.32\textwidth,height=105pt]{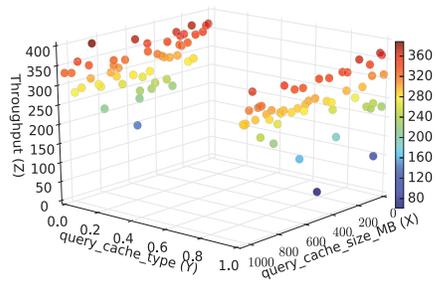}\vspace{-6pt}
      \caption{The performance surface for MySQL under the \textbf{Zipfian} read-write workload. (Best view in color)}\vspace{-12pt}
      \label{fig:mysqlrw} 
\end{figure}

Configuration tuning is crucial to obtaining a good performance from a deployed system. For an application workload, a configuration setting leads to the best performance of the system, but it might not be optimal given another application workload. Take Figure~\ref{fig:mysql} for example. Under the \emph{uniform-read} workload, the value of \emph{query\_cache\_type} is key to a good performance; but, as shown in Figure~\ref{fig:mysqlrw}, the \emph{query\_cache\_type} value has no obvious relation with the system performance for a \emph{Zipfian read-write} workload. In fact, the default setting generally cannot achieve the best performance of a system under all workloads.

Configuration tuning is highly time-consuming and laborious. It requires the users: (1) to find the heuristics for tuning; (2) to manually change the system configuration settings and run workload tests; and, (3) to iteratively go through the second step many times till a satisfactory performance is achieved. Sometimes, the heuristics in the first step might be misguiding, as some heuristics are correct for one workload but not others; then, the latter two steps are in vain. In our experience of tuning MySQL, it has once taken five junior employees about half a year to find an appropriate configuration setting for our cloud application workloads.

Configuration tuning is even not easy for experienced developers. For example, it has been shown that, although PostgreSQL's performance under the workload of a TPC-H query is a smooth surface with regard to the configuration parameters of cache size and buffer size~\cite{ituned}, the cache size interacts with the buffer size in a way that even takes the system developers great efforts to reason about the underlying interactions. In fact, the system performance models can be highly irregular and complicated, as demonstrated by Figure~\ref{fig:mysql} to \ref{fig:spark} and Figure~\ref{fig:mysqlrw}. How the configuration settings can influence the system performance can hardly be anticipated by general users or expressed by simple models.

{\large \textbf{Benefits}}. Automatic configuration tuning can greatly benefit system users. First, a good configuration setting will improve the system performance by a large margin, while automatic configuration tuning can help users find the good configuration setting. Second, a good configuration setting is even more important for repetitive workloads, and recurring workloads are in fact a common phenomenon~\cite{repeatWork1,repeatWork2}. Third, automatic configuration tuning enables fairer and more useful benchmarking results, if the system under test is automatically tuned for a best configuration setting before benchmarking---because the system performance is related to both the workload and the configuration setting.
\subsection{Challenges}

Several challenges exist for automatic configuration tuning for general systems. These challenges must be addressed \emph{simultaneously}.

\textbf{Variety of performance goals:} Users can have different performance goals for configuration tuning. For a data analytical job on Spark, the performance goal is normally to reduce the running time, while for a data-access workload on MySQL, it would be to increase the throughput. Sometimes, users can have multiple performance goals, e.g., increasing the throughput and decreasing the average latency of individual operations for MySQL. Some users would also require to improve the performance goal such as throughput but not to worsen other metrics such as memory usage. Besides, some performance goals need to be maximized, while some need to be minimized.

\textbf{Variety of systems and workloads:} To tune the variety of systems and workloads, we cannot build or have users build performance models for the tuning purpose as previously done for Hadoop~\cite{starfish}. Some deployed systems are distributed, e.g., Spark and Hadoop, while some are standalone, e.g., Tomcat or one-node Hadoop. A system's performance model can be strongly influenced by the hardware and software settings of the deployment environment~\cite{acts}. Hence, the automatic configuration tuning system must handle the variety of deployment environments. It must enable an easy usage with the deployed systems and workloads. The heterogeneity of deployed systems and workloads can have various performance models for tuning, leading to different best configuration settings and invalidating the reuse of samples across different deployments~\cite{acts,ottertune}.

\textbf{High-dimensional parameter space:} As mentioned previously, many systems in the cloud now have a large number of configuration parameters, i.e., a high-dimensional parameter space for configuration tuning. On the one hand, it is impossible to get the complete image of the performance-configuration relations without samples covering the whole parameter space. On the other hand, collecting too many samples is too costly. The typical solutions to the optimization problem over high-dimensional spaces generally assume the abundance of samples. For example, some solve the optimization problem with around \emph{10} parameters using about \emph{2000} samples~\cite{rrs,paraTuning}. Except through simulations, it is too costly to collect such an amount of samples in practice; thus, such solutions are not applicable to the configuration tuning problem of real systems.

\textbf{Limited samples:} It is impractical to collect a large number of performance-configuration samples in practice. Besides, it is impossible to build a performance simulator for every system in the cloud, thus making the simulation-based sample collection infeasible. We have to collect samples through real tests against the deployed systems. Thus, methods used for configuration tuning cannot rely on a large sample set. Rather, it should produce results even on a limited number of samples. And, as the number of samples is increased, the result should be improved.
\begin{figure*}[!t]
      \centering
      \includegraphics[width=0.95\textwidth]{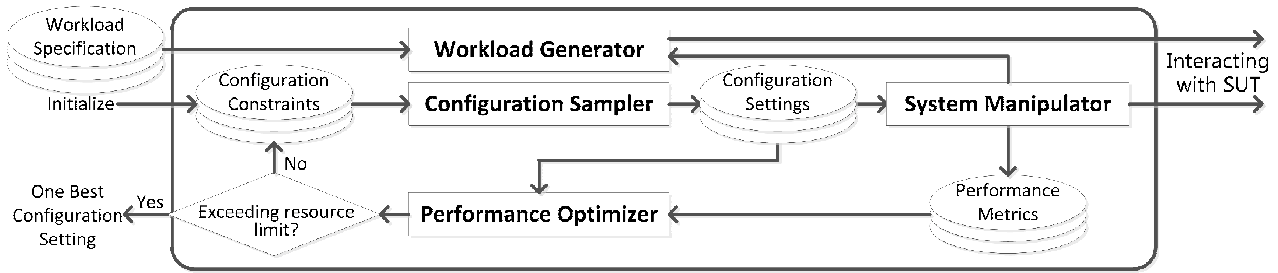}\vspace{-6pt}
      \caption{The automatic configuration tuning process and the major components of BestConfig.}\vspace{-3pt}
      \label{fig:process} 
\end{figure*}
\section{BestConfig Design}
\label{sec:main}

BestConfig is designed to automatically find, within a given resource limit, a configuration setting that can optimize the performance of a deployed system under a specific application workload. We call the process of adjusting configuration settings as configuration tuning (or just tuning) and the system to adjust as SUT (System Under Tune).

\subsection{Design Overview}

To satisfy users' various needs on performance optimization and simultaneously simplify the optimization problem, we adopt the utility function approach to amalgamating multiple performance optimization needs into a single maximization goal. BestConfig exposes an interface for users to express their performance optimization goals ($\S$\ref{sec:goal}).

To handle the variety of deployed systems and workloads, BestConfig is designed with a flexible architecture that has loosely coupled components, as sketched in Figure~\ref{fig:process}. These components are connected through data flows. The system manipulator component is an interface to interact with an SUT deployed in the target environment, while the workload generator component allows the easy plug-in of any target workload.

With limited samples, the tuning process must collect samples by careful choices. Different combinations of deployed systems and workloads can require different sampling choices. Thus, we design the BestConfig architecture with a sampler component that interacts with the system manipulator on sample collection. Besides, the performance optimization process can introduce more information on which configuration settings to sample; thus, the performance optimizer component of BestConfig is designed to interact with the sampler to pass on such knowledge. The resulting architecture of BestConfig is detailed in Section~\ref{sec:arch}.

To address the configuration tuning problem, we must solve the two subproblems of \textbf{sampling} and \textbf{performance optimization (PO)} simultaneously. Due to the challenges of high-dimensional parameter space and limited samples, the sampling subproblem differs from the random sampling in related works. It must be solved with additional conditions as detailed in Section~\ref{sec:subproblems}. The PO subproblem also faces similar conditions ($\S$\ref{sec:subproblems}). A feasible solution to automatic configuration tuning for general systems must address all the conditions for the two subproblems.

BestConfig exploits the sampling information when solving the PO subproblem, and vice versa. In contrast, sampling and performance optimization are generally addressed \textbf{separately} in related works. We combine the sampling method DDS (Divide and Diverge Sampling) with the optimization algorithm RBS (Recursive Bound and Search) as a complete solution. DDS will sample for later RBS rounds in subspaces that are not considered in early RBS rounds. In this way, the requirement on wide space coverage for sampling is better satisfied. Furthermore, RBS exploits DDS in the bounded local search to reduce the randomness and increase the effectiveness of the search. In comparison, sampling was rarely considered and exploited for the local search step in related works. We detail DDS and RBS in Section~\ref{sec:ddsrbs}. In the following, we first describe the key steps for automatic configuration tuning ($\S$\ref{sec:steps}).

\subsection{Key Steps for Configuration Tuning}%
\label{sec:steps}

Figure~\ref{fig:process} sketches the automatic configuration tuning process of BestConfig. The tuning process is in closed loop. It can run in as many loops as allowed by the resource limit. The resource limit is typically the number of tests that are allowed to run in the tuning process. It is provided as an input to the tuning process. Other inputs include the configuration parameter set and their lower/upper bounds (denoted as \emph{configuration constraints}). The output of the process is a configuration setting with the optimal performance found within a given resource limit.

Given \emph{configuration constraints}, the configuration sampler generates a number of configuration settings as allowed by the resource limit. The configuration settings are then used to update the configuration setting of the SUT. For each configuration setting, a test is run against the SUT; and, the corresponding performance results are collected. The performance results are then transformed into a scalar performance metric through the utility function.

All the sample pairs of the performance metric and the corresponding configuration setting are used by the performance optimization (PO) algorithm. The PO algorithm finds a configuration setting with the best performance. If the resource limit permits more tests and samples, the PO algorithm will record the found configuration setting and output a new set of configuration constraints for the next tuning loop. Otherwise, the tuning process ends and BestConfig outputs the configuration setting with the best performance found so far.\vspace{-9pt}

\subsection{Performance Metric by {\large Utility Function}}%
\label{sec:goal}

BestConfig optimizes towards a scalar performance metric, which has only a single value. The scalar performance metric is defined by a \emph{utility function}, with user-concerned performance goals as inputs. If only one performance goal is concerned, e.g., the throughput or the latency, the utility function is the identity function, i.e., $f(x)=x$, where $x$ is the performance goal. If multiple performance goals are concerned simultaneously, we can define the utility function as a weighted summation. For example, if a user wants to increase the throughput and decrease the latency, the utility function can be defined as $f(x_t,x_l)=x_t/x_l$, where $x_t$ is the throughput and $x_l$ the latency. In case that the throughput must be increased and the memory usage must not exceed the threshold $c_m$, an example utility function is $f(x_t,x_m)=x_t\times S(c_m-x_m-5)$, where $x_m$ is the memory usage and $S(x)$ is the sigmoid function $S(x)=\frac{1}{1+e^{-x}}$. BestConfig allows users to define and implement their own utility functions through a \texttt{Performance} interface~\cite{bestconf}.

During the configuration tuning process, BestConfig maximizes the performance metric defined by the utility function. Although users can have performance goals that need to be minimized, we can easily transform the minimization problem into a maximization one, e.g., taking the inverse of the performance metric.\vspace{-3pt}

\subsection{Highly-Extensible Architecture}%
\label{sec:arch}

BestConfig has a highly flexible and extensible architecture. The architecture implements the flexible configuration tuning process in closed loop. It allows BestConfig to be easily used with different deployed systems and workloads, requiring only minor changes.

The main components in BestConfig's architecture include \emph{Configuration Sampler}, \emph{Performance Optimizer}, \emph{System Manipulator} and \emph{Workload Generator}. \emph{Configuration Sampler} implements the scalable sampling methods. \emph{Performance Optimizer} implements the scalable optimization algorithms. \emph{System Manipulator} is responsible for updating the SUT's configuration setting, monitoring states of the SUT and tests, manipulating the SUT, etc. \emph{Workload Generator} generates application workloads. It can be a benchmark system like YCSB~\cite{ycsb} or BigOP~\cite{ycsb} running a benchmarking workload; or, it can be a user-provided testing system regenerating the real application workloads. The system manipulator and the workload generator are the only two components interacting with the SUT.

For extensibility, the components in the architecture are loosely coupled. They only interact with each other through the data flow of configuration constraints, configuration settings and performance metrics. The configuration sampler inputs the system manipulator with sets of configuration settings to be sampled. The system manipulator inputs the performance optimizer with the samples of performance-configuration pairs. The performance optimizer adaptively inputs new configuration constraints to the configuration sampler. With such a design, BestConfig's architecture allows different scalable sampling methods and scalable PO algorithms to be plugged into the configuration tuning process. On coping with different SUTs or workloads, only the system manipulator and the workload generator need to be adapted. With the extensible architecture, BestConfig can even optimize systems emerging in the future, with only slight changes to the system manipulator and the workload generator.\vspace{-3pt}

\subsection{An Example of Extending BestConfig}

With the current Java implementation of BestConfig~\cite{bestconf}, one can define a new sampling method by implementing the {\tt ConfigSampler} interface. The {\tt ConfigSampler} interface accepts the sample set size limit and a list of configuration constraints as inputs, and returns a list of configuration settings.

Similarly, to plug in a new PO algorithm, one can implement the {\tt Optimization} interface. The implementation of the {\tt Optimization} interface can accept a list of configuration settings and their corresponding performance metrics from the system manipulator. It must decide whether to continue the automatic configuration process or not, based on the given resource limit, e.g., the number of tests allowed. The best configuration setting can be output to a file, while the new configuration constraints are directly passed to the configuration sampler.

At present, extending the system manipulator requires only changing a few shell scripts that interact with the SUT and the workload generator. The workload generator is loosely coupled with other system components. Thus, it is highly convenient to integrate user-provided workload generation systems, e.g., YCSB and HiBench bundled in the BestConfig source~\cite{bestconf}. Thanks to the highly extensible architecture, we have already applied BestConfig to six systems as listed in Table~\ref{tbl:paraNums}. The corresponding shell scripts for these systems are provided along with the BestConfig source.\vspace{-6pt}

\subsection{Subproblems: Sampling and PO}%
\label{sec:subproblems}

\textbf{The subproblem of sampling} must handle all types of parameters, including boolean, enumeration and numeric. The resulting samples must have a wide coverage of the parameter space. To guarantee resource scalability, the sampling method must also guarantee a better coverage of the whole parameter space if users allow more tuning tests to be run. Thus, the sampling method must produce sample sets satisfying the following three conditions: (1) the set has a wide coverage over the high-dimensional space of configuration parameters; (2) the set is small enough to meet the resource limit and to reduce test costs; and, (3) the set can be scaled to have a wider coverage, if the resource limit is expanded.

\textbf{The subproblem of performance optimization (PO)} is to maximize the performance metric based on the given number of samples. It is required that the output configuration setting must improve the system performance than a given configuration setting, which can be the default one or one manually tuned by users. To optimize the output of a function/system, the PO algorithm must satisfy the following conditions: (1) it can find an answer even with a limited set of samples; (2) it can find a better answer if a larger set of samples is provided; and, (3) it will not be stuck in local sub-optimal areas and has the possibility to find the global optimum, given enough resources. Two categories of PO algorithms exist, i.e., model-based~\cite{starfish,tkdeConfTune,ottertune} and search-based~\cite{rrs,smarthillclimbing,paraTuning}. In the design of BestConfig, we exploit the search-based methods.

We do not consider model-based PO methods for the following reasons. First, with the large number of configuration parameters, model-based methods would require a large number of samples to construct a useful model, thus violating the first condition of the PO subproblem. Second, model-based methods require the user to have a priori knowledge about the model, e.g., whether the model should be linear or quadratic, but it is mission impossible for general users to input such a priori information for each combination of SUT, deployment setting and workload. Third, model-based methods have hyper-parameters, which strongly impact how the model works; but setting these hyper-parameters is as hard as tuning the configuration setting of the SUT. Without enough samples, precise a priori knowledge or carefully-tuned hyper-parameters, model-based methods will not even work. In Section~\ref{sec:model}, we experiment with two model-based methods using limited samples. We demonstrate that these model-based methods hardly work in the configuration tuning problem with a resource limit.\vspace{-6pt}

\section{DDS \& RBS in Cooperation}%
\label{sec:ddsrbs}

To address the two subproblems of automatic configuration tuning, we propose the divide-and-diverge sampling (DDS) method and the recursive bound-and-search (RBS) algorithm. Although the sampling and PO methods can work separately, DDS and RBS in cooperation enables the effective tuning process of BestConfig.

\subsection{DDS: Divide \& Diverge Sampling}%
\label{sec:DDS}

\textbf{Parameter space coverage}. To guarantee a wide coverage over the high-dimensional parameter space, we \textbf{divide} the space into subspaces. Then we can randomly select one point from each subspace. Thus, each subspace is represented by one sample. In comparison to the random sampling without subspace division, it is very likely that some subspaces are not represented, especially when the dimension of the space is high. Given $n$ parameters, we can divide the range of each parameter into $k$ intervals and collect combinations of the intervals. There are $k^n$ combinations, thus $k^n$ subspaces and samples. This way of sampling is called \emph{gridding} or \emph{stratified sampling}. Thanks to subspace division, gridding guarantees a complete coverage of the whole parameter space. But it also results in a sample set with a large cardinality, which is in exponential relation to the number of parameter dimensions. Hence, it violates the second requirement of the sampling subproblem.

\textbf{Resource limit}. To meet the second requirement, we reduce the number of subspaces to be sampled. We observe that, \textbf{the impact of an influential parameter's values on the performance can be demonstrated through comparisons of performances, disregard of other parameters' values}. For example, consider the performance model of MySQL as plotted in Figure~\ref{fig:mysql}. If the value of a parameter has great impacts on the performance like $query\_cache\_type$, we actually do not need to examine all combinations of the parameter's values with every other parameter's values. Instead, we need only examine each potentially outstanding value of the parameter once and compare the resulting performance with other samples. Thus, given a limited resource, we consider each interval of a parameter once, rather than making full combinations of all intervals.

\begin{figure}[!t]
      \centering
      \includegraphics[width=0.33\textwidth]{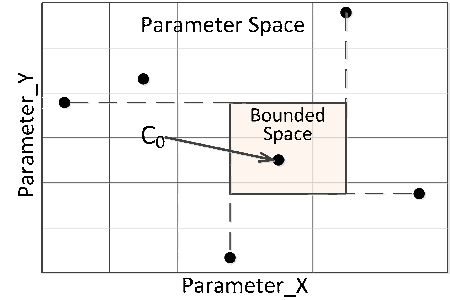}\vspace{-6pt}
      \caption{An example of running DDS and RBS for a 2D space.}\vspace{-12pt}
      \label{fig:dsprocess} 
\end{figure}

After dividing parameter ranges into $k$ intervals, we do not make a full combination of all intervals. Rather, we take a permutation of intervals for each parameter; then, we align the interval permutation for each paremeter and get $k$ samples. For example, with two parameters $X$ and $Y$ divided into $6$ range intervals respectively, we can take $6$ samples as demonstrated in Figure~\ref{fig:dsprocess}. Each range interval of $X$ is represented exactly once by the sample set. So is that of $Y$. For a given sample-set size, we \textbf{diverge} the set of sample points the most by representing each interval of each parameter exactly once.

\textbf{Scalability}. The third requirement for sampling is to be scalable with regard to the resource limit, e.g., the number of tests allowed, while meeting the previous two requirements. In fact, the above \textbf{divide-and-diverge sampling} (DDS) method directly meets the third requirement. The value of $k$ is set according to the resource limit, e.g., $k$ being equal to the number of tests allowed. Increasing the number of allowed tests, the number of samples will increase equally; thus, the parameter space will be divided more finely and the space coverage will be increased.

Furthermore, as the configuration tuning process is in closed loop, multiple times of sampling can be run. For the sake of scalability and coverage, DDS do not complete restart a new sampling process by redividing the whole space. Rather, on a request of resampling, DDS reuses its initial division of the whole parameter space and samples in subspaces not considered previously, while diverging the sample points as much as possible.

{\large \textbf{Heterogeneity of parameters}}. Although DDS considers the continuous range of a parameter, DDS can be applied to parameters of boolean or categorical types by transforming them into parameters with continuous numeric ranges. Take the boolean type for example. We can first represent the parameter value of \emph{true} and \emph{false} by $1$ and $0$ respectively. Then, we let the values be taken from the range of $[0,2)$, to which the DDS method can be directly applied. We can map a sampled value within ranges of $[0,1)$ and $[1,2)$ respectively to the values of $0$ or $1$, which are equal to \emph{false} and \emph{true} respectively. Similar mappings can be carried out for categorical or enumerative parameters as well.
\subsection{RBS: Recursive Bound \& Search}%
\label{sec:RBS}

Consider the performance surfaces in Figure~\ref{fig:tomcat} and Figure~\ref{fig:spark}. These performance plots have parameters with numeric values and continuous ranges, thus the performance surfaces are continuous surfaces. \textbf{Given a continuous surface, there is a high possibility that we find other points with similar or better performances around the point with the best performance in the sample set.} Even if the continuous performance surface might not be smooth, e.g., that in Figure~\ref{fig:tomcat}, or if the performance surface is continuous only when projected to certain dimensions, e.g., that in Figure~\ref{fig:mysql} when constrained to specific subspaces, the above observation still applies. Based on this observation, we design the RBS (Recursive Bound and Search) optimization algorithm.

\textbf{Bound step}. Given an initial sample set, RBS finds the point $C_0$ with the best performance. Then, it asks for another set of points sampled in the \emph{bounded space} around $C_0$. Based on the observation in the last paragraph, there is a high possibility that we will find another point (say $C_1$) with a better performance. We can again sample in a bounded space around $C_1$. We can recursively carry out this bound-and-sample step until we find no point with a better performance in a sample set.

Here, there is a problem of how large the \emph{bounded space} should be. According to the observation in Section~\ref{sec:DDS}, if a performance value shall have influential and positive impacts on the performance, it shall lead to a high performance in the sample set. For the initial sample set, parameter values other than those represented by $C_0$ are actually not having positive impacts as influential as $C_0$ on the performance, thus we should not consider them again given the limited resource. In other words, the \emph{bounded space} around $C_0$ shall not include parameter values represented by any other points in the sample set. In addition, a high performance might be achieved by any parameter value of those around $C_0$ but unrepresented in the sample set.

RBS fixes the bounds of the \emph{bounded space} as follows. For each parameter $p_i$, RBS finds the largest value $p_i^f$ that is represented in the sample set and that is smaller than that of $C_0$. It also finds the smallest value $p_i^c$ that is represented in the sample set and that is larger than that of $C_0$. For the dimension represented by the parameter $p_i$, the bounded space has the bounds of $(p_i^f,p_i^c)$. Figure~\ref{fig:dsprocess} demonstrates this bounding mechanism of RBS. The same bounding mechanism can be carried out for every $C_j,j=0,1,...$ in each bound-and-sample step.

By now, RBS has addressed the first two requirements for the PO subproblem. It finds an answer even with a limited set of samples by recursively taking the bound-and-sample step around the point $C_j$, which is the point with the best performance in a sample set. Let each bound-and-sample step called a \emph{round}. RBS can adjust the size of the sample set and the number of rounds to meet the resource limit requirement. For example, given a limit of $nr$ tests, RBS can run in $r$ rounds with each sample set sized $n$. Given more resources, i.e., a larger number of allowed tests, RBS can carry out more bound-and-sample steps to search more finely in promising bounded subspaces.

\textbf{Recursion step}. To address the third requirement and avoid being stuck in a sub-optimal bounded subspace, RBS restarts from the beginning of the search by having the sampler to sample in the complete parameter space, if no point with a better performance can be found in a bound-and-sample step. This measure also enables RBS to find a better answer if a larger set of samples is provided. This is made possible through searching around more promising points scattered in the huge high-dimensional parameter space.\vspace{3pt}

\subsection{Why Combining {\large DDS} with {\large RBS} Works}%
\label{sec:reason}

In this section, we discuss about why combing DDS with RBS works in the configuration tuning problem. The performance of a system can be represented by a measurable objective function $f(x)$ on a parameter space $D$. In DDS, $D$ is divided into orthogonal subspaces $D_{i}$. We define the distribution function of objective function values as:
\begin{equation}
\phi_{D_{i}}(y_0) = \frac{m(\{x\in D_{i}| f(x)\leq y_0\})}{m(D)}
\end{equation}
where $y_0$ is the performance for the default configuration setting $P_0$ and $m(\cdot)$ denotes \emph{Lebesgue measure}, a measure of the size of a set. For example, \emph{Lebesgue measure} is area for a set of 2-dimensional points, and volume for a set of 3-dimensional points, and so on. The above equation thus represents the portion of points that have no greater performance values than $P_0$ in the subspace. The values of $\phi_{D_{i}}(y_0)$ fall within the range of $[0,1]$. If a subspace has no points with greater performance values than $y_0$, it will have a zero value of $\phi(y_0)$. When all points in a subspace have higher performances, the subspace will have $\phi(y_0)$ evaluated to one.

DDS divides the whole high-dimensional space into subspaces and then samples in each subspace. A sample can be either greater or no greater than the default performance $y_0$. Assume all points with no greater performances are in set $s_{i0}$ and those with greater ones are in set $s_{i1}$, we have $m(s_{i})=m(s_{i0})+m(s_{i1})$. Given $\phi_{D_{i}}(y_0)$ for subspace $D_{i}$, randomly sampling according to the \emph{uniform distribution} will result in a $\phi_{D_{i}}(y_0)$ probability of getting points with no better performances and a $1-\phi_{D_{i}}(y_0)$ probability of getting points with better performances.

Randomly sampling according to the uniform distribution, DDS will output samples with greater performances after around $n=1/(1-\phi_{D_{i}}(y_0))$ samples for subspace $D_{i}$. Although the exact value of $n$ is not known, the principle underlying the uniform-random number generation guarantees that more samples will finally lead to the answer. In other words, given enough resources (i.e., samples), DDS will get a point with a greater performance than $P_0$.

RBS bounds and samples around the point with the best performance in a sample set. This key step works because, if a point $C_j$ in a subspace $D_{i}$ is found with the best performance, it is highly probable that the subspace $D_{i}$ has a larger value of $1-\phi_{D_{i}}(y_0)$ than the other subspaces, as all subspaces are sampled for the same number of times in all rounds of RBS. According to the definition of $\phi_{D_{i}}(y_0)$, subspaces with larger values of $1-\phi_{D_{i}}(y_0)$ shall have more points that lead to performances greater than $y_0$, as compared to subspaces with smaller values of $1-\phi_{D_{i}}(y_0)$. Thus, RBS can scale down locally around $C_j$ to search again for points with better performances. As a result, the bound step of RBS, recursively used with DDS, will lead to a high probability of finding the point with the optimal performance. If the small probability event happens that the bound step runs in a subspace with a relatively small value of $1-\phi_{D_{i}}(y_0)$, the phenomenon of trapping in the local sub-optimal areas occurs. The recursion step of RBS is designed to handle this situation by sampling in the whole parameter space again.\vspace{3pt}

\section{Evaluation}%
\label{sec:eval}

We evaluate BestConfig on six widely deployed systems, namely Hadoop~\cite{hadoop}, Hive~\cite{hive}, Spark~\cite{spark}, Cassandra~\cite{cassandra}, Tomcat~\cite{tomcat}, and MySQL~\cite{mysql}. These systems are deployed for Huawei's applications named Cloud+ and BI. To generate workloads towards systems under tune, we embed widely adopted benchmark tools in the workload generator. We use HiBench~\cite{hibench} for Hive+Hadoop and Spark, YCSB~\cite{ycsb} for Cassandra, SysBench~\cite{sysbench} for MySQL and JMeter~\cite{jmeter} for Tomcat. Table~\ref{tbl:paraNums} summarizes the evaluated systems along with the corresponding numbers of tuned parameters respectively. The detailed lists of the tuned parameters, as well as the detailed descriptions of the SUTs and the evaluated workloads, are accessible on the Web~\cite{bestconf}.
\begin{table}[!b]
  \centering
  \vspace{-6pt}
  \caption{The evaluated systems and parameters.}\vspace{-6pt}%
  \label{tbl:paraNums}%
  \begin{tabular}{lllc}
\toprule[1.2pt]
  \multirow{2}*{\small \textbf{Software}} & \multirow{2}*{\small \textbf{Description}} & \multirow{2}*{\small \textbf{Language}} & \textbf{\small \# Parameters}\\
  & & & \textbf{\small Tuned}\\
  \midrule[0.8pt]
  \textbf{Spark} & Distributed computing & Scala & \textbf{30}\\
  \midrule[0.2pt]
  \textbf{Hadoop} & Distributed computing  & Java & \multirow{2}*{\textbf{109}}\\
  \textbf{Hive} & Data analytics  & Java & {\small (in all)}\\
  \midrule[0.2pt]
  \textbf{Cassandra} & NoSQL database & Java & \textbf{28}\\
  \midrule[0.2pt]
  \textbf{MySQL} & Database server  & C++ & \textbf{11}\\
  \midrule[0.2pt]
  \textbf{Tomcat} & Web server & Java & \textbf{13} \\
\bottomrule[1.2pt]
\end{tabular}
\end{table}

Our experimental setup involves multiple local clusters of servers to deploy the six systems. If not specifically mentioned, the server is equipped with two 1.6GHz processors that have two physical cores, and 32GB memory, running CentOs 6.0 and Java 1.7.0\_55. To avoid interference and comply with the actual deployment, we run the system under tune, the workload generator and other components of BestConfig on different servers. Further details can be found on the Web~\cite{bestconf}.

In the evaluation, we answer five questions:\vspace{-2pt}
\begin{enumerate}
  \item Why the configuration tuning problem with a resource limit is nontrivial ($\S$\ref{sec:model});\vspace{2pt}
  \item How well BestConfig can optimize the performance of SUTs ($\S$\ref{sec:evalTune});\vspace{2pt}
  \item How effective the cooperation of DDS and RBS is ($\S$\ref{sec:evalDDS});\vspace{2pt}
  \item How the sample set size and the number of rounds affect the tuning process ($\S$\ref{sec:evalSizeRound});\vspace{2pt}
  \item Whether the configuration setting found by BestConfig will maintain its advantage over the given setting in tests outside the tuning process ($\S$\ref{sec:evalWorkloads}).
\end{enumerate}
\begin{figure*}[t]
\centering
 \begin{minipage}{0.43\textwidth}
      \centering
     \includegraphics[width=\textwidth]{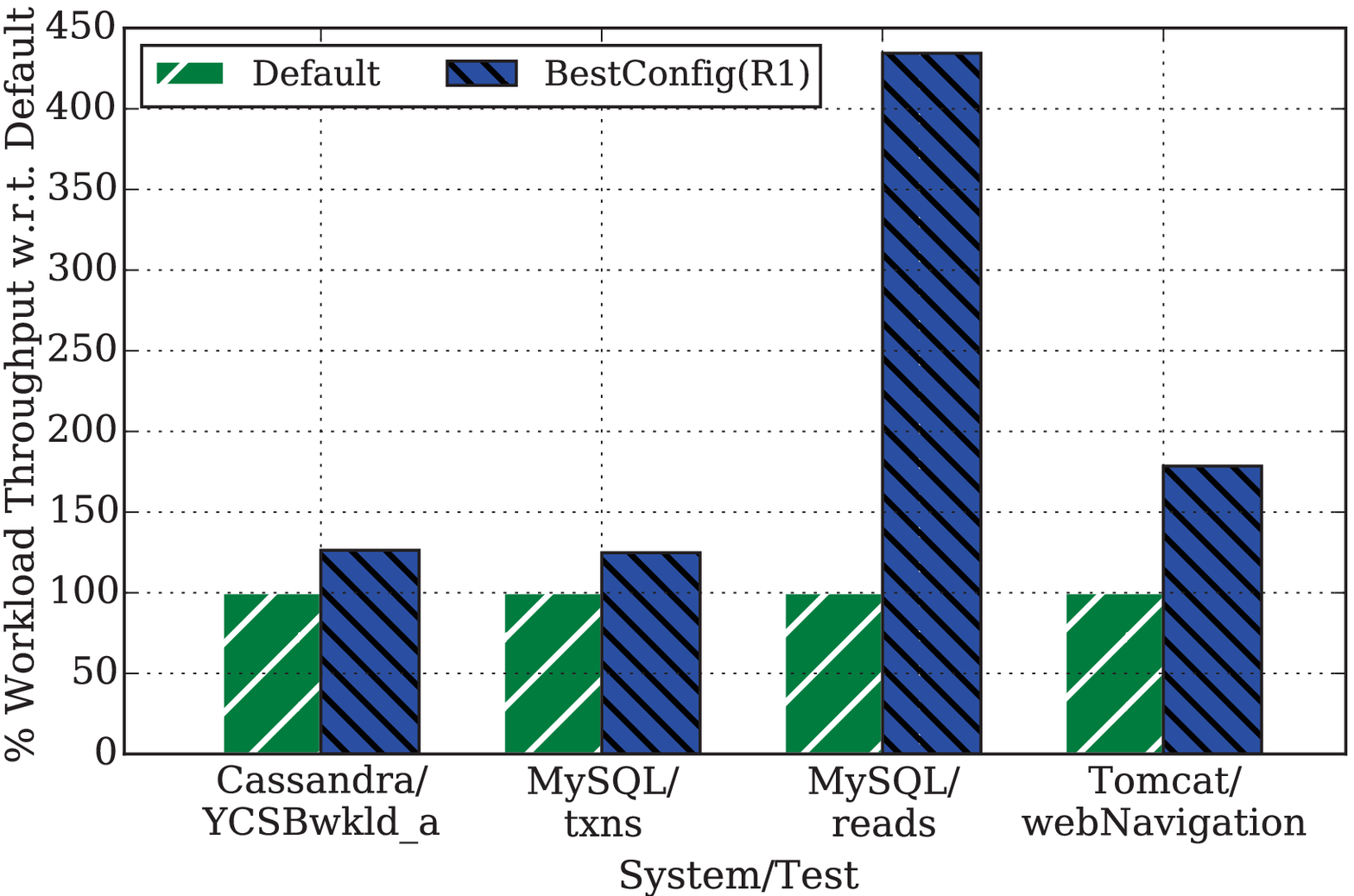}\vspace{-6pt}
      \label{fig:qualityThroughput} 
  \end{minipage}\vspace{-6pt}
 \begin{minipage}{0.43\textwidth}
     \centering
     \includegraphics[width=\textwidth]{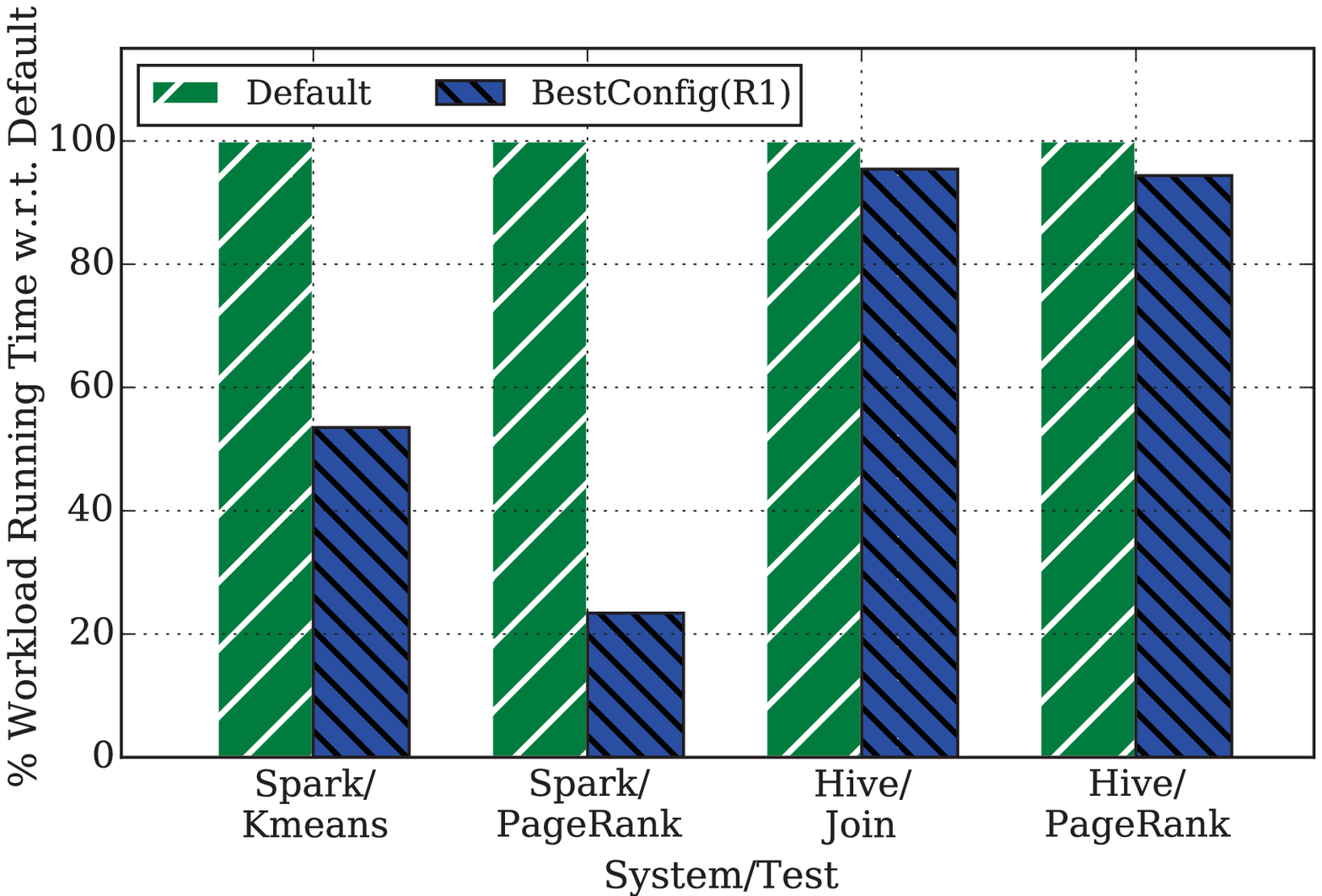}\vspace{-6pt}
      \label{fig:qualityDuration} 
    \end{minipage}\vspace{-6pt}
    \caption{BestConfig's optimization capability with regard to the default configuration setting.}\vspace{-12pt}
    \label{fig:result}
\end{figure*}
\subsection{Infeasibility of Model-based Methods}%
\label{sec:model}

The difficulty of the configuration tuning problem can be demonstrated by the infeasibility of model-based methods. Common machine learning methods are model-based methods. They were previously used in optimization problems that have only a limited number of parameters. Based on the highly extensible architecture of BestConfig, we implement two PO algorithms, adopting the machine learning approach.

One is based on the COMT (Co-Training Model Tree) method~\cite{comt}, which assumes a linear relation between parameters and the performance. COMT divides the parameter space into subspaces and builds linear models for each subspace. Therefore, many common linear models can be taken as special cases of COMT. Besides, COMT is a semi-supervised machine learning method, which is designed and expected to work with a limited number of samples.

We train the COMT model using training sets of 100, 200 and 300 samples respectively. Each training set is randomly selected from a pool of 4000 samples, which are generated in a BestConfig tuning experiment over the Tomcat deployment described above. According to the COMT algorithm, the training not only exploits the training set, but also another set of unsampled points to reduce generalization errors. We validate the three learned models on the testing set with all samples in the sample pool. We summarize the prediction errors in Table~\ref{tbl:comt}, where \emph{Avg. err. rate} is the average error rate and \emph{Max. err. rate} the maximum error rate. Here, \emph{error rate} is computed as the actual performance dividing the difference between the predicted performance and the actual performance.
\begin{table}[!h]
\vspace{-2pt}
  \centering
  \small
  \caption{Linear-model based performance predictions.}\vspace{-6pt}%
  \label{tbl:comt}%
  \begin{tabular}{ccc}
\toprule[1.1pt]
 { \textbf{Sample set size}} & { \textbf{Avg. err. rate}} & { \textbf{Max. err. rate}}\\
  \midrule[0.8pt]
    100 &  14\%	& 240\% \\
      \midrule[0.2pt]
    200&  15\%	&  1498\% \\
      \midrule[0.2pt]
    300&  138\% & 271510\% \\
\bottomrule[1.1pt]
\end{tabular}
\end{table}

From Table~\ref{tbl:comt}, we can see that the predictions are in fact very much inaccurate. Although the first two average error rates look small, the corresponding models can make highly deviated predictions. The reason that more samples lead to worse predictions is twofold. One is because of model overfitting, and the other is due to the highly irregular performance surface of the SUT.

The other machine learning model we have tried is the GPR (Gaussian Process Regression) method~\cite{ituned}, which assumes a differentiable performance function on parameters. It is the state-of-the-art model-based method adopted in a recent work on database tuning~\cite{ottertune}. GPR does not predict the performance for a given point. Rather, it constructs the model based on the covariances between sample points and outputs points that are most probably to increase the performance the most, i.e., to achieve the best performance.

We experiment GPR using training sets with 100, 200 and 300 samples respectively. These sample sets are also collected from BestConfig tuning experiments over the Tomcat deployment described above. Among all the provided samples, GPR make a guess on which point would lead to the best performance (\emph{best guess}). We then run a test on the best-guess point to get the actual performance. We compare the actual performance for GPR's best guess with that for the default configuration setting (\emph{default}). Besides, we compare GPR's best guess with the real best point that has the optimal performance among all the provided samples, denoted as \emph{real best}. The results are given in Table~\ref{tbl:gpr}. We can see that, although the prediction is improving as the number of samples increases, GPR's predictions about best points are hardly accurate.
\begin{table}[!h]
\vspace{-6pt}
  \centering
  \small
  \caption{GPR-based predictions on best points.}\vspace{-6pt}%
  \label{tbl:gpr}%
  \begin{tabular}{ccc}
\toprule[1.1pt]
 { \textbf{Sample set size}} & { \textbf{Bst. guess/dflt.}} & { \textbf{Bst. guess/rl. bst.}}\\
  \midrule[0.8pt]
    100 &  93\%	& 56\% \\
      \midrule[0.2pt]
    200&  104\%	& 63\% \\
      \midrule[0.2pt]
    300&  121\% & 58\% \\
\bottomrule[1.1pt]
\end{tabular}\vspace{-6pt}
\end{table}

Because of the complexity of performance models, common model-based optimization methods, e.g. COMT and GPR, do not work well in the configuration tuning problem. In essence, the assumptions of such algorithms do not hold for the SUTs. As a result, methods like COMT and GPR cannot output competitive configuration settings. Moreover, their results do not guarantee to improve as the number of samples is increased, i.e., not scalable with the resource limit; instead, their results might worsen because of overfitting, violating the conditions of the PO subproblem.

\subsection{Automatic Configuration Tuning Results}%
\label{sec:evalTune}

Figure~\ref{fig:result} presents the automatic configuration tuning results for Cassandra, MySQL, Tomcat, Spark and Hive+Hadoop using BestConfig. For the latter three systems, we ran two tuning experiments with different benchmark workloads on each system. In all experiments, we set the sample set size to be 100 and the round number to be one. As demonstrated by the results, BestConfig improves the system performances in all experiments. Even though the Hive+Hadoop system has 109 parameters to tune, BestConfig can still make a performance gain. In comparison to the other SUTs, the performance gain for Hive+Hadoop is relatively small. The underlying reason is that this SUT has almost 10 times as many configuration parameters as the other SUTs.

\begin{figure}[!b]
      \centering
	\vspace{-12pt}
      \includegraphics[width=0.48\textwidth]{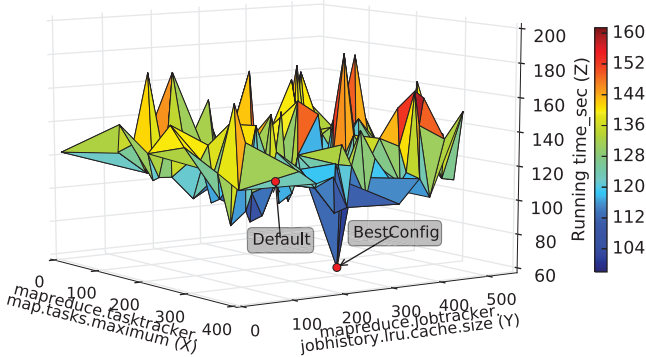}\vspace{-6pt}
      \caption{BestConfig reduces 50\% running time of HiBench-Join on Hive+Hadoop within 500 tests.}
      \label{fig:hiveSurf} 
\end{figure}

However, \textbf{BestConfig can improve the tuning result as the size of the sample set is increased}. Setting the sample set size to be 500, we carry out another experiment of Hive+Hadoop under the HiBench Join workload. The result is demonstrated in Figure~\ref{fig:hiveSurf}. The BestConfig setting reduces 50\% running time of the Join job.

To sum up the results, BestConfig has improved the throughput of Tomcat by about 75\%, that of Cassandra by about 25\%, that of MySQL by about 430\%, and reduced the running time of Hive join job by about 50\% and that of Spark join job by about 80\%, solely by configuration adjustments.

\begin{figure}[!b]
      \centering
      \vspace{-12pt}
      \includegraphics[width=0.45\textwidth]{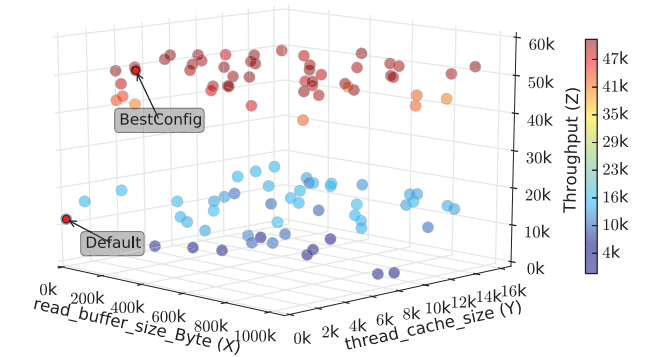}\vspace{-6pt}
      \caption{Throughputs for varied \emph{thread\_cache\_size} of MySQL, invalidating the manual tuning guideline.}
      \label{fig:mysqlScatter} 
\end{figure}

{\large \textbf{Invalidating manual tuning guidelines}}. The results produced by BestConfig have invalidated some manual tuning rules. For example, some guideline for manually tuning MySQL says that the value of \emph{thread\_cache\_size} should never be larger than 200. However, according to BestConfig's results demonstrated in Figure~\ref{fig:mysqlScatter}, we can set the parameter to the large value of $11987$, yet we get a much better performance than following the guideline.
\subsection{Cooperation of DDS and RBS}%
\label{sec:evalDDS}

\begin{figure}[!t]
      \centering
      \includegraphics[width=0.435\textwidth]{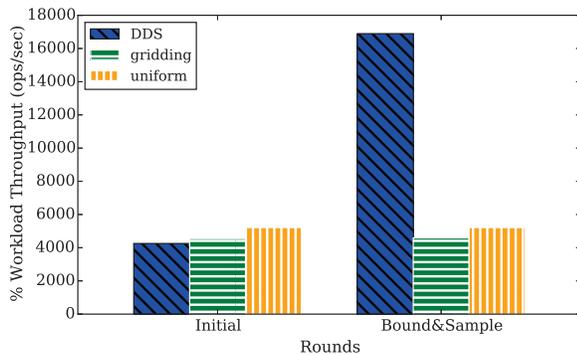}\vspace{-12pt}
      \caption{Sampling method comparisons: DDS outperforms gridding and uniform in the latter round.}\vspace{-15pt}
      \label{fig:dds} 
\end{figure}

We have evaluated the DDS (divide-and-diverge sampling) method of BestConfig as compared to uniform random sampling and gridding sampling. We carry out the comparisons based on Tomcat through tuning two configuration parameters. In the experiments, we use all sampling methods with RBS. We set the initial sample set size to be 100 and the round number to be 2. Thus, after sampling for the first round, each sampling method will sample around a promising point in the second round, denoted as \emph{bound and sample}. The results are plotted in Figure~\ref{fig:dds}.

In the initial round, the three sampling methods have sampled points with similar best performances. The effectiveness and advantage of DDS is demonstrated in the bound-and-sample round. As shown in Figure~\ref{fig:dds}, DDS have sampled points with the best performance as much as three times more than those of the gridding and the uniform sampling. The advantage of DDS over the gridding is due to its diverging step, while that over the uniform sampling is due to the complete coverage of the sampling space. DDS considers 100 diversities for each parameter, while the gridding considers only 10. And, there is a possibility that the uniform sampling will take samples locating at some restricted area of the space, while DDS is guaranteed to scatter samples across the space and with divergence.
\begin{figure}[!b]
      \centering
      \vspace{-12pt}
      \includegraphics[width=0.39\textwidth]{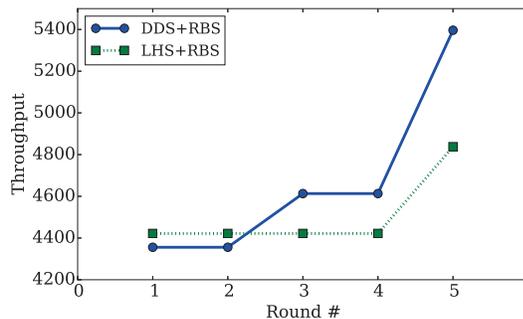}\vspace{-12pt}
      \caption{DDS+RBS makes progress in earlier rounds than LHS+RBS.}
      \label{fig:ddsvlhs} 
\end{figure}

We also compare DDS with LHS (Latin Hypercube Sampling)~\cite{lhs}. LHS can produce the same sample sets as DDS in one-time sampling. However, DDS differs from LHS in that DDS remembers previously sampled subspaces and resamples towards a wider coverage of the whole parameter space. This difference leads to the DDS method's advantage of coverage and scalability over LHS. This advantage is demonstrated in Figure~\ref{fig:ddsvlhs} through a configuration tuning process for Tomcat. In this tuning process, we set the sample set size for each round as 100 (according to the experimental results of Table~\ref{tbl:sizeRound}). We can see that DDS in cooperation with RBS makes progress in earlier rounds than LHS with RBS.
\begin{figure}[!t]
      \centering
      \includegraphics[width=0.42\textwidth]{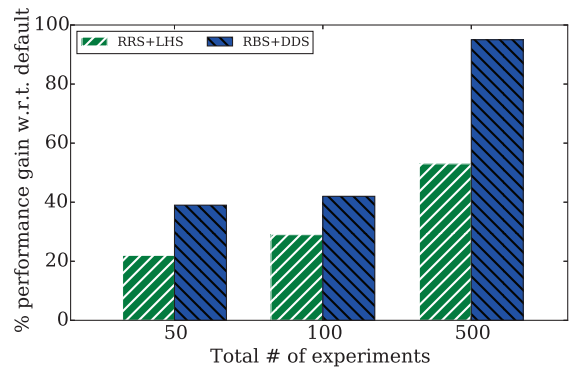}\vspace{-12pt}
      \caption{RBS+DDS vs. RRS+LHS.}\vspace{-15pt}
      \label{fig:rbsVrrs} 
\end{figure}

Furthermore, we try replacing RBS with RRS (Recursive Random Search)~\cite{rrs}. RRS is a search-based optimization method with the exploitation and exploration steps, similar to the bound and recursion steps of RBS. However, RBS are designed with space coverage and scalability, while RRS has no such preferred properties. Hence, when we compare RBS+DDS with RRS+LHS in experiments on a Tomcat deployment, the former outperforms the latter given the same resource limit. The results are demonstrated in Figure~\ref{fig:rbsVrrs}.

\subsection{Varied Sample-Set Sizes \& Rounds}%
\label{sec:evalSizeRound}

To understand how the sample set size and the number of rounds affect the optimization process, we limit the number of tests to 100 and carry out five sets of experiments with varied sample-set sizes and rounds. We vary the sample-set sizes from 5 to 100 and the number of rounds from 20 to 1 accordingly. The experiments are run upon Tomcat using the webpage navigation workload, tuning 13 parameters.

The first five rows of Table~\ref{tbl:sizeRound} summarizes the results for each set of experiments, regarding the performance gains for the initial sampling-search round and the whole tuning process. As the size of the sample set increases, both performance gains are increasing. This fact implies that DDS is scalable. In the meantime, given a limited resource, we should set a sample-set size as large as possible, before we increase the number of rounds.

However, \textbf{a larger sample-set size for one round does not necessarily always indicate a better tuning result.} We have experimented with 500 samples for one round, tuning the Tomcat deployment. We find that little performance gain is obtained over the tuning process with 100 samples for one round, as demonstrated by the last row of Table~\ref{tbl:sizeRound}. In comparison, the tuning process of Figure~\ref{fig:hiveSurf}, which also uses 500 samples for one round, makes much more performance gain than when using 100 samples for one round. The reason behind the difference lies in \textbf{the number of parameters}. In our experiments, Tomcat has only 13 parameters, while Hive+Hadoop has 109. The more parameters, the larger sample-set size is required.
\begin{table}[!b]
  \centering
  \vspace{-12pt}
  \caption{Performance gains on varied sample-set sizes and rounds.}\vspace{-6pt}%
  \label{tbl:sizeRound}%
  \begin{tabular}{ccc}
\toprule[1.2pt]
 { \textbf{Exps (SetSize*rounds)}} & { \textbf{Initial gain}} & { \textbf{Best gain}}\\
  \midrule[0.8pt]
    5*20& $\mathbf{0\%}$	&$\mathbf{5\%}$\\
      \midrule[0.2pt]
    10*10&  $\mathbf{0\%}$	&$\mathbf{14\%}$\\
      \midrule[0.2pt]
    20*5& $\mathbf{26\%}$	&$\mathbf{29\%}$\\
      \midrule[0.2pt]
    50*2& $\mathbf{39\%}$	&$\mathbf{39\%}$\\
      \midrule[0.2pt]
    100*1& $\mathbf{42\%}$	&$\mathbf{42\%}$\\
     \midrule[0.2pt]
    500*1& $\mathbf{43\%}$	&$\mathbf{43\%}$\\
\bottomrule[1.2pt]
\end{tabular}
\end{table}

\textbf{Rounds matter}. Despite that a large initial sample-set size is important, more rounds are necessary for better tuning results. Consider Figure~\ref{fig:ddsvlhs} again. Because of randomness, it is not guaranteed that more rounds mean definitely better results. For example, the second round of DDS+RBS does not actually produce a better result than the first round. However, more rounds can lead to better results, e.g., the third round and the fifth round of DDS+RBS in Figure~\ref{fig:ddsvlhs}. In fact, the third round and the fifth round are executing the recursion step of RBS. This step is key to avoiding suboptimal results. Thus, by searching in the whole parameter space again, the third and fifth rounds find configuration settings with higher performances. How much BestConfig can improve the performance of a deployed system depends on factors like SUT, deployment setting, workload and configuration parameter set. But BestConfig can usually tune a system better when given more resource and running more rounds than when given less resource and running fewer rounds.
\begin{figure}[!t]
      \centering
      \includegraphics[width=0.45\textwidth]{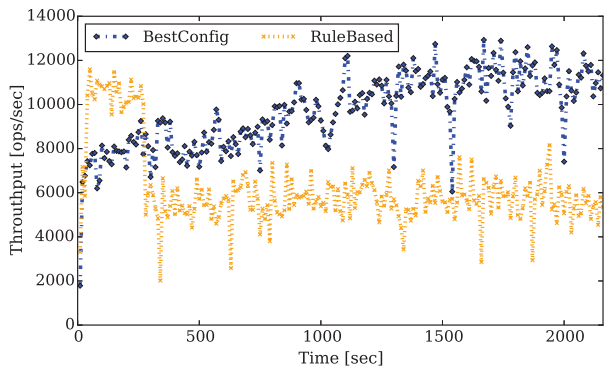}\vspace{-9pt}
      \caption{A long-running test of the manually-tuned and BestConfig settings for Cassandra under Huawei's Cloud+ application workloads.}\vspace{-12pt}
      \label{fig:stable} 
\end{figure}
\subsection{Stable Advantage of {\large BestConfig} Setting}%
\label{sec:evalWorkloads}

BestConfig generally adopts short tests in the tuning process. For the long running application workloads, it might seem that the iterative testing and configuration tuning process cannot work. We argue that, even though the automatic configuration tuning process for such workloads might be long, the performance gained from the long tuning process is still worthwhile. Besides, as the benchmarking community has proved theoretically and practically~\cite{JimAnon,tpchistory,tpc,spec,berkeleyview,bigop}, the long-running application workloads can be represented by some short-running workloads.

As demonstrated by our experience with an actual BestConfig deployment, short tests can represent long-running workloads, if the tests are specified and generated properly.  We have deployed BestConfig to tune the Cassandra system for Huawei's Cloud+ applications. For confidentiality reasons, we simulated the application workloads using the YCSB benchmark, which is then integrated to the workload generator of BestConfig. In the automatic configuration tuning process, the simulated workload is run for about ten minutes. We set the sample set size to be 60 and the round number as one. As output by BestConfig, a configuration setting was found with about $29\%$ performance gain than the setting tuned by Huawei engineers. Later, we applied the configuration setting found by BestConfig to Cassandra and ran the workload for about 40 minutes. A similar long-running test is carried out with the Huawei-tuned configuration setting as well. The resulting throughput is demonstrated in Figure~\ref{fig:stable}.

As shown in Figure~\ref{fig:stable}, BestConfig's configuration setting keeps its advantage over the one set by Huawei engineers. In fact, BestConfig's setting has an average throughput of 9679 ops/sec, while Huawei's rule-based setting can only achieve one of 5933 ops/sec. Thus, BestConfig has actually made a $63\%$ performance gain. This performance improvement is made merely through configuration adjustments.

In fact, BestConfig can usually tune a system to a better performance than the manual tuning that follows common guidelines recommended for general system deployments. The reasons are twofold. First, such manual tuning might achieve a good performance for many system deployments under many workloads, but the tuned setting is usually not the best for a specific combination of SUT, workload and deployment environment. As demonstrated in Section~\ref{sec:evalTune}, some tuning guidelines might work for some situations but not the others. Second, the number of configuration parameters is too large and the interactions within a deployed system are too complex to be comprehended by human~\cite{asilomar}.\vspace{-3pt}
%
%
\subsection{Discussion}

Users ought to specify a resource limit in proportion to the number of parameters for tuning. Although BestConfig can improve the performance of a system based on a small number of samples, there is a minimum requirement on the number of samples. Consider Table~\ref{tbl:paraNums}. If the user allows only 5 samples in a round for tuning 13 parameters, the user is not likely to get a good result. When the number of samples exceeds that of parameters, e.g., from the second row of Table~\ref{tbl:paraNums}, the performance of the system gets improved obviously. Similarly, for a system with more than 100 parameters to tune, BestConfig can only improve the system performance by about 5\%, if only 100 samples are provided (Figure~\ref{fig:result}). However, when the sample set size is increased to 500, BestConfig can improve the system performance by about 50\% (Figure~\ref{fig:hiveSurf}).

If we can reduce the number of parameters to tune, we can reduce the number of tuning tests and fasten the tuning process, since the number of parameters is related to the number of samples needed for tuning. A recent related work on configuration tuning has proposed to reduce the number of parameters through a popular linear-regression-based feature selection technique called Lasso~\cite{ottertune}. We consider integrating similar parameter reduction methods into BestConfig as future work.

BestConfig can generally do a great job if given the whole set of parameters. Even if the set of parameters is not complete, BestConfig can generally improve the system performance as long as the set contains some parameters affecting the system performance. In case that the set of parameters to tune are totally unrelated to an SUT's performance, BestConfig will not be able to improve the SUT's performance. Besides, BestConfig cannot improve an SUT's performance if (1) the SUT is co-deployed with other systems, which are not tuned by BestConfig and which involve a performance bottleneck affecting the SUT's performance; or, (2) the SUT with the default configuration setting is already at its optimal performance.

However, if the above situations occur, it means that the SUT's performance cannot be improved merely through configuration adjustments. Instead, other measures must be taken such as adding influential parameters for tuning, removing bottleneck components or improving the system design.\vspace{-6pt}

\section{Use Case: Tomcat for Cloud+}%
\label{sec:case}

BestConfig has been deployed to tune Tomcat servers for Huawei's Cloud+ applications. The Tomcat servers run on virtual machines, which run on physical machines equipped with ARM CPUs. Each virtual machine is configured to run with 8 cores, among which four are assigned to process the network communications. Under the default configuration setting, the utilizations of the four cores serving network communications are fully loaded, while the utilizations of the other four processing cores are about 80\%. With such CPU behaviors, Huawei engineers have considered that the current throughput of the system is the upper bound and no more improvement is possible.

Using BestConfig and setting the overall throughput as the performance metric for optimization, we then found a configuration setting that can improve the performance of the deployment by 4\%, while the CPU utilizations remain the same. Later, the stability tests demonstrate that the BestConfig setting can guarantee the performance improvement stably. The results of the stability tests are demonstrated in Table~\ref{tbl:tomcatOnArm}. We can observe improvements on every performance metric by using the BestConfig setting.

Thus, BestConfig has made it possible to improve the performance of a fully loaded system by simply adjusting its configuration setting. It has expanded our understanding on the deployed systems through automatic configuration tuning.
\begin{table}[!t]
  \centering
  \caption{BestConfig improving performances of a fully-loaded Tomcat.}\vspace{-6pt}%
  \label{tbl:tomcatOnArm}%
  \begin{tabular}{lllr}
\toprule[1.2pt]
 { \textbf{Metrics}} & { \textbf{Default}} & { \textbf{BestConfig}} & \textbf{Improvement}\\
  \midrule[0.8pt]
    Txns/seconds& 978 & 1018 & $\mathbf{4.07\%\uparrow}$\\
      \midrule[0.2pt]
    Hits/seconds& 3235 & 3620 &  $\mathbf{11.91\%\uparrow}$\\
      \midrule[0.2pt]
    Passed Txns& 3184598& 3381644&  $\mathbf{6.19\%\uparrow}$\\
      \midrule[0.2pt]
    Failed Txns& 165& 144&  $\mathbf{12.73\%\downarrow}$\\
      \midrule[0.2pt]
    Errors& 37& 34&  $\mathbf{8.11\%\downarrow}$\\
\bottomrule[1.2pt]
\end{tabular}\vspace{-12pt}
\end{table}
\section{Related Work}%
\label{sec:related}

The closest related works for BestConfig are the classic Latin Hypercube Sampling (LHS) method~\cite{lhs} and the recursive random search (RRS) algorithm~\cite{rrs}. DDS differs from LHS in that DDS remembers previously sampled subspaces and resamples towards a wider coverage of the whole parameter space. This difference leads to the DDS method's advantage of coverage and scalability over LHS. RBS differs from RRS in two aspects. First, RRS requires the users to set multiple hyper-parameters, which have strong impacts on the optimization results, but setting hyper-parameters is as hard as configuration tuning. Second, RRS searches a local subspace by examining one sample after another. Such a design is efficient only if the local search is limited to a small space, but this is generally not true for high-dimensional spaces. If the hyper-parameters are carefully set as for a narrow local search, then the space not examined would be too large. Such trade-off is difficult. Besides, searching a local space by taking one sample at a time involves too much randomness; in comparison, the local search of BestConfig takes advantage of the sampling method. One more crucial difference is that BestConfig exploits RBS and DDS together.

Quite a few past works have been devoted to automatic configuration tuning for Web systems. These works either choose a small number of parameters to tune, e.g., smart hill climbing~\cite{smarthillclimbing}, or require a huge number of initial testings~\cite{eurosysConf,paraTuning}, e.g., simulated annealing~\cite{sa} and genetic algorithms~\cite{ga}. Although constructing performance models might help finding appropriate configuration settings~\cite{qog}, a large number of samples will be required for modeling in a large configuration parameter space. But collecting a large set of samples requires to test the SUT for many times. This is a highly costly process. A related work uses reinforcement learning in the same tuning problem~\cite{RLweb}. It formulates the performance optimization process as a finite Markov decision process (MDP), which consists of a set of states and several actions for each state. The actions are increasing or decreasing the values of individual parameters. As mentioned previously, the performance functions of deployed systems can be complicated, e.g., with many sudden ups and downs on the surface. A seemingly wise step with some performance gain might result in a bad final setting, due to the local suboptimal problem.

Works on automatic configuration tuning for database systems also exist. iTuned~\cite{ituned} assumes a smooth performance surface for the SUT so as to employ the Gaussian process regression (GPR) for automatic configuration tuning. But the assumption can be inapplicable to other SUTs, e.g., Tomcat or MySQL given some specific set of configuration parameters. The recent work of OtterTune~\cite{ottertune} also exploits GPR. It additionally introduces a feature-selection step to reduce the number of parameters. This step reduces the complexity of the configuration tuning problem. We are examining  the possibility of integrating similar feature selection methods into BestConfig to reduce the number of configuration parameters before starting the tuning process.

Automatic configuration tuning is also proposed for the Hadoop system. Starfish~\cite{starfish} is built based upon a strong understanding of the Hadoop system and performance tuning. Thus, the method used in Starfish cannot be directly applied to other systems. Aloja~\cite{aloja} adopts the common machine learning methods, exploiting a large database of samples. But, samples are costly to obtain. As analyzed in Sectio~\ref{sec:subproblems} and \ref{sec:model}, Aloja's approach is not applicable to the configuration tuning of general systems. 


\section{Conclusion}%
\label{sec:conclude}

We have presented the automatic configuration tuning system BestConfig. BestConfig can automatically find, within a given resource limit, a configuration setting that can optimize the performance of a deployed system under a specific application workload. It is designed with a highly flexible and extensible architecture, the scalable sampling method DDS and the scalable performance optimization algorithm RBS. As an open-source package, BestConfig is available for developers to use and extend in order to effectively tune cloud systems. We have used BestConfig to tune the configuration settings of six widely used systems and observed the obvious performance improvements after tuning. Furthermore, tuning the Tomcat system on virtual machines in the Huawei cloud, BestConfig has actually made it possible to improve the performance of a fully loaded system by simply adjusting its configuration settings. These results highlight the importance of an automatic configuration tuning system for tapping the performance potential of systems.

\begin{acks}
We would like to thank our shepherd, Ennan Zhai, and the anonymous reviewers for their constructive comments and inputs to improve our paper. We would like to thank the Huawei Cloud+ and the Huawei BI teams in helping us verify BestConfig towards their online applications. This work is in part supported by the National Natural Science Foundation of China (Grant No. 61303054 and No. 61420106013), the State Key Development Program for Basic Research of China (Grant No. 2014CB340402) and gifts from Huawei.
\end{acks}

\balance

\bibliographystyle{ACM-Reference-Format}
\bibliography{ref}

\end{document}